\documentclass[fleqn,10pt]{wlscirep}
\usepackage[utf8]{inputenc}
\usepackage[T1]{fontenc}
\usepackage{graphicx, caption, subcaption}
\usepackage{enumitem}
\usepackage{float}
\usepackage{ulem}

\usepackage{color}

\renewcommand{\emph}{\textit}
\renewcommand{\em}{\it}

\title{Stability of the personal relationship networks in a longitudinal study of middle school students}

\author[1]{Diego Escribano}
\author[2,3]{Francisco J. Lapuente}
\author[1,4]{Jos\'e A. Cuesta}
\author[5]{Robin I.M. Dunbar}
\author[1,4,*]{Angel S\'anchez}

\affil[1]{Grupo Interdisciplinar de Sistemas Complejos (GISC), Departamento de
Matem\'aticas, Universidad Carlos III de Madrid, 28911 Legan\'es, Madrid,
Spain}
\affil[2]{Instituto de Enseñanza Secundaria Blas de Otero, 28024 Madrid, Spain}
\affil[3]{Departamento de Biología y Geología, Física y Química Inorgánica, Universidad Rey Juan Carlos, 28933, Móstoles, Madrid, Spain}
\affil[4]{Instituto de Biocomputaci\'on y F\'\i sica de Sistemas Complejos (BIFI), Universidad de Zaragoza, 50018 Zaragoza, Spain}
\affil[5]{Department of Experimental Psychology, University of Oxford, Oxford OX2 6GG, UK}
\affil[*]{Corresponding author; e-mail: anxo@math.uc3m.es}

\begin{abstract}
{ The personal network of relationships is structured in circles of friendships, that go from the most intense relationships to the least intense ones. While this is a well established result, little is known about the stability of those circles and their evolution in time.} To shed light on this issue, we study the temporal evolution of friendships among teenagers during two consecutive academic years by means of a survey administered on five occasions.  We show that the first two circles, best friends and friends, can be clearly observed in the survey but also that being in one or the other leads to more or less stable relationships.  We find that being in the same class is one of the key drivers of friendship evolution. We also observe an almost constant degree of reciprocity in the relationships, around 60\%, a percentage influenced both by being in the same class and by gender homophily. Not only do our results confirm the mounting evidence supporting the circle structure of human social networks, but they also show that these structures persist in time despite the turnover of individual relationships---a fact that may prove particularly useful for understanding the social environment in middle schools.
\end{abstract}
\begin{document}

\flushbottom
\maketitle
\thispagestyle{empty}
\section*{Introduction}

Human egocentric social networks have very distinct sizes and structures. In adults, they comprise around 150 friends and family members, organised in a hierarchically inclusive series of circles of about 5, 15, 50  and, finally, the 150, with each circle being approximately three times the size of the circle immediately inside it.\cite{dunbar2020} The sizes of these circles appear to be remarkably stable across time, despite changes in membership.\cite{saramaki2014} This is largely a consequence of the fact that the time available for socialising (and hence the capacity for bond-building) is strictly limited.\cite{roberts2009,roberts2011,miritello2013} The circles are created by differential time investment, with the five members of the innermost circle receiving around 40\% of total social effort between them. The second layer, i.e., the additional 10 members of the 15-circle, receive an additional 20\% of the time investment, with the 135 members of the outer network layer sharing the remaining 40\%.\cite{sutcliffe2012} The circles themselves appear as the most likely distribution of time under this overall constraint in a context where the benefits a relationship provides are proportional to the time invested in it.\cite{tamarit2018,escribano2021} { The fact that circles are a consequence of the limitations in the total effort available can been found in other, non-human primates having a similar structure.\cite{escribano2022b}}

{While the circle structure of personal networks is well established, much less is known about its evolution in time. A great deal of analysis, whose meta-analysis is presented in Ref.\ \citen{wrzus2013}, has focussed on how the size and composition of personal networks change over time, showing, for instance, that they increase until adulthood and then decrease, and that life events modulate them. Data analysis of a massive set of mobile phone calls \cite{kunal2016} confirms these results and reveals the role played by gender in the life cycle of friendships. Other approaches have considered the small timescale evolution of networks, using data from Bluetooth, phone calls and social media with temporal resolutions as short as five minutes \cite{sekara2016}, although the focus of this study was the social network as a whole rather than personal networks. However, none of these studies have taken the perspective of the circle structure to explore the time evolution of relationships.}

{In this paper, we study the evolution of personal networks placing our emphasis on the circle structure and an intermediate time scale. Thus, we have collected data from teenagers by using surveys administered every three to five months during two years (each of these surveys will be referred to as a ``wave''), allowing us to capture the evolution of the personal networks for a significant period of time. The main reason for choosing this population is its availability, as the students were enrolled in the same school for the whole period of the study and we could survey them repeatedly. In this line of research,}  it is worth mentioning recent results by Kucharski {\em et al.} \cite{kucharski2018}, who show that a longitudinal approach yields consistent results, in the sense that the properties of data remain essentially the same across successive waves. As we will show below, our results confirm this consistency. Other longitudinal studies have shown, for instance, that students prefer to gradually reorganise their social networks according to their performance levels\cite{smirnov2017} or that residential and school mobility has a large impact on the structure of adolescents' friendship networks.\cite{south2004} Studies of these age groups have focused on tie quality in a handful of identified close family and friend relationships (‘best friends’ or, perhaps, the inner core of five best friends). There is some evidence that network size increases across childhood and the teenage years, and that it does so by adding whole layers in step-changes rather than by accreting alters individually.\cite{roberts2011}. 

Here we report longitudinal data on the inner circles of friendship networks in young teenagers. We focus on the 5- and 15-circles among a cohort of middle school children (corresponding to years 1 to 4 of mandatory middle school (``enseñanza secundaria obligatoria'', ESO in Spanish), and how these change both over time and in the face of the constraints imposed by lockdown during COVID. {Students turn 12 the year they begin 1 ESO, and should turn 16 the year they finish 4 ESO, although a few of them are a year older because they had to do the same course twice (called in Spanish ``repetidores''). Each year is in turn divided into classes, and the class structure was different in the two academic years under consideration (2020-2021 and 2021-2022) due to COVID prevention measures (see Methods for a detailed description). In the text, when we refer to a ``class'' we mean the administrative unit, even if they are of different sizes in 1 and 2 ESO in the two academic years or if in one year they take turns to be at school. We consider the effect of a number of variables on their relationships: course (as a proxy for age), class (because students sharing classes interact much more than outside their classes), gender (it is well-known that there are strong gender homophily effects in teenagers, particulary in the age range we are considering\cite{shrum88}), itinerary (because students following their lessons in English or Spanish tend to interact only with those in their same itinerary) and ``repetidor'' (because not being promoted to the next level means you do not share class with your former friends anymore).} We study two key aspects of friendship networks:  their size and composition and the stability of these relationships as students change classes in successive years and are exposed to new social opportunities. { We report the results of five waves of our social relationship survey, and we will refer to them by their number. The surveys were carried out on the following dates: December 2020 (wave 1), May 2021 (wave 2), October 2021 (wave 3), February 2022 (wave 4), and May 2022 (wave 5). This means that waves 1 and 2 correspond to the academic year 2020-2021 and waves 3-5 correspond to the academic year 2021-2022. Academic years in middle schools in Spain begin around mid-September and end around mid-June.} We note that due to COVID the change in the academic year between wave 2 and wave 3 students involved a reshuffling of the classes, whose effects will also be of interest for our study.
{ We obtain information on both positive relationships associated with the second (just friends) and the first (best friends) circles and on negative relationships. While a detailed study of enmities is beyond the scope of the present work, we collected these data because they provide a potentially different light on friendships as we will see below. In addition, it allows us to see whether friendship terminations are due to castastrophic falling out or just simple decay.
}
The results reported here largely extend those presented in Ref.~\citen{escribano2021}, which were limited to only 1 ESO and two waves, carried out in December 2018 and May 2019. See Methods for more details on the sample, the school organization during the COVID pandemic in the course 2020-2021, and the outlier detection procedure.

\section*{Results}

\begin{figure}
\includegraphics[width=0.975\textwidth]{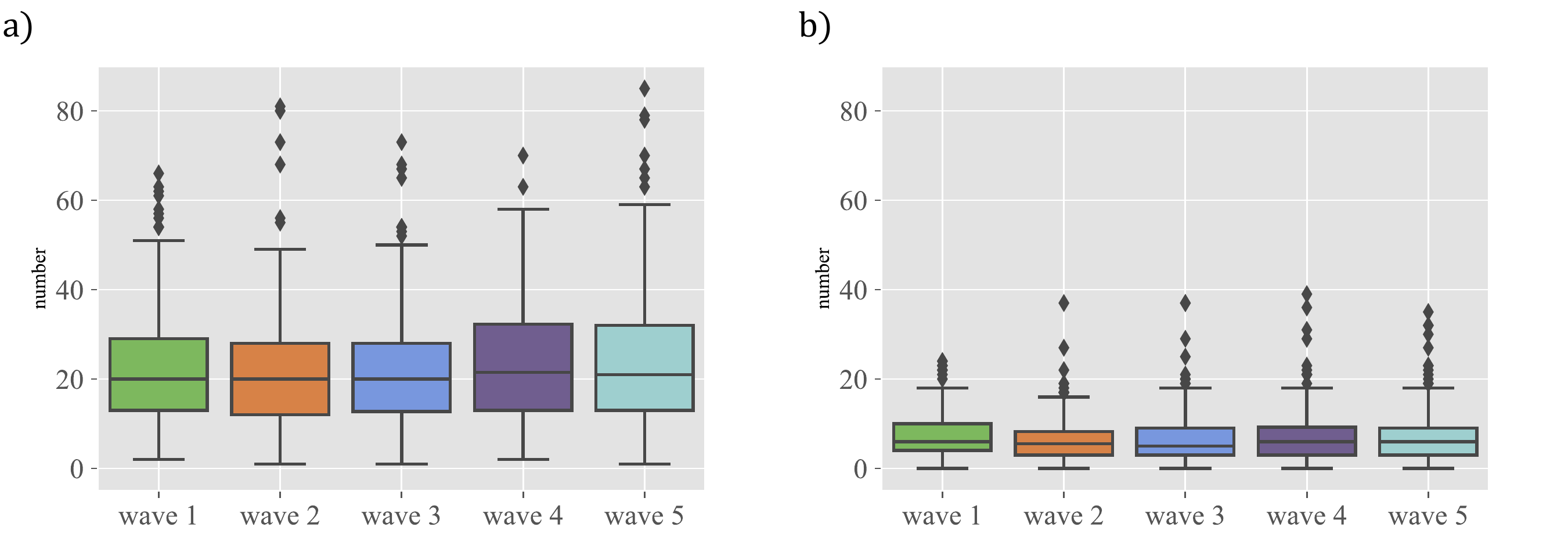}
\caption{{\textbf{The structure of friendships} - The figure represents  boxplots corresponding to: a) the number of total friendships (best friendships + just friendships) and b) the number of best friendships reported by the students in each of the five waves of the survey. }}
\label{fig_relationships}
\end{figure}

The number of friendships and best friendships the students have is remarkably consistent across the five waves despite the very different situations in the two academic years under study (cf.~Fig.~\ref{fig_relationships}), in agreement with Ref.~\citen{kucharski2018}. Even with the drastic upheaval of COVID in waves 4 and 5, the results are very similar across all the waves. The results are the same even if splitting the data by class, gender, itinerary, or by ``repetidor''  (see plots in section S2 in the Supplementary Information). This is true of both just friendships and best friendships. If a regression line is set through the number of friends in each of the five waves for each student, the mean of that slope is exactly 0 (see Supplementary Information, Sec.~S4). 

To delve deeper into the structure of ego-networks, we employed the parameter $\mu$ as a tool for analysing the circular structure of a given individual, introduced in Ref.~\citen{tamarit2018} and discussed in Ref.~\citen{escribano2021}. This parameter is obtained for every individual by fitting the analytical expression from Ref.~\citen{tamarit2018} to the reported numbers of friendships in the two circles. When $\mu$ is greater than 0, the circles have a characteristic structure where the number of friendships grows rapidly as we move away from circle 1. A value of $\mu$ close to 0.7 is typically seen when the scaling ratio between the sizes of the circles is around 3, as is often observed. Conversely, when $\mu$ is less than 0, most friendships are concentrated in circle 1 and the additional circles have very few additional people. These negative values of $\mu$ are observed in situations where the number of possible links is limited (e.g.~sailors in a boat, communities of migrants, etc.\cite{tamarit2018}), or for introverted individuals. The distribution of the $\mu$ parameters (cf.~Ref.~\citen{escribano2021}) is also constant and, in fact, most individual values of $\mu$ change very little in the five waves (see plots in section S3 in the Supplementary Information; see also the distribution of the slopes of fits to the evolution of $\mu$ in Sec. S4) in spite of the organisational changes due to COVID and to the composition of the classes when moving to the next academic year (see Methods).

In order to study the evolution of relationships over time, we have compared the nature of links (labelled $+2$ for `best friend', $+1$ for `just friend', $0$ for `no link', $-1$ for `just enemy', and $-2$ for `worst enemy') between pairs of subjects in two consecutive waves ($w_n$ and $w_{n+1}$), and have computed diverse conditional probabilities. Thus, Fig.~\ref{fig_circles} represents $P(x,w_5|+2,w_4)$ {(a)}  and $P(x,w_5|+1,w_4)$ {(b)}, whereas Fig.~\ref{fig_circles2} represents $P(x,w_5|-2,w_4)$ {(a) and $P(x,w_5|-1,w_4)$ {(b)}. (Other conditional probabilities are shown in Sec.~S5 of the Supplementary Information.)

The first two circles (best friends and friends, respectively) evolve in very different ways: best friends are quite stable and when they stop being best friends they end up as just friends most of the time. In contrast, friends are more dynamic and may disappear from the radar or, in some cases, even become best friends. 
We have also looked at the opposite evolution, namely where students that appear for the first time as best friends were in previous waves (i.e.~$P(x,w_{n-1}|+2,w_n)$). (All these conditional probabilities are shown in Sec.~S5 of the Supplementary Information.) We have observed that best friends were often already best friends in the previous wave, and new ones come mostly from being friends. Therefore, the first two circles show clear differences in the stability of the relationships they involve, arising from the different intensity in best friends vs just friends relationships.

{ Even if it is not the main objective of our research, we find it interesting to report on our results on negative relationships, or enmities, a subject that has been largely overlooked in the past to the extent that very few results are available. In fact, in the context of schools we are only aware of the contemporaneous work by S\'anchez-Espinosa {\em et al} \cite{sanchez23}, while in a more general (historical) context we can mention the work of Yose {\em et al} \cite{kenna2017}.}
Enmities are very few and highly volatile. The total number of negative relationships is an order of magnitude smaller than that of positive ones. As can be seen from the plot (Fig.~\ref{fig_circles2}, and Sec. S4 in Supplementary Information for the rest of the waves), most relationships that are marked as worst enemies or just enemies in one wave end up not marked in the next wave. Interestingly, worst enemies are retained with higher frequency than just enemies. These results point to friendships and enmities having a different nature, with friendships being more long-lasting and enmities reflecting, in general, more the heat of a specific conflict---although very bad relationships may last longer. { The lack of a significant number of enmities prevents us from doing a deeper study, but these first insights suggest that it would be a very relevant one. }

\begin{figure}
\centering
\includegraphics[width=0.975\textwidth]{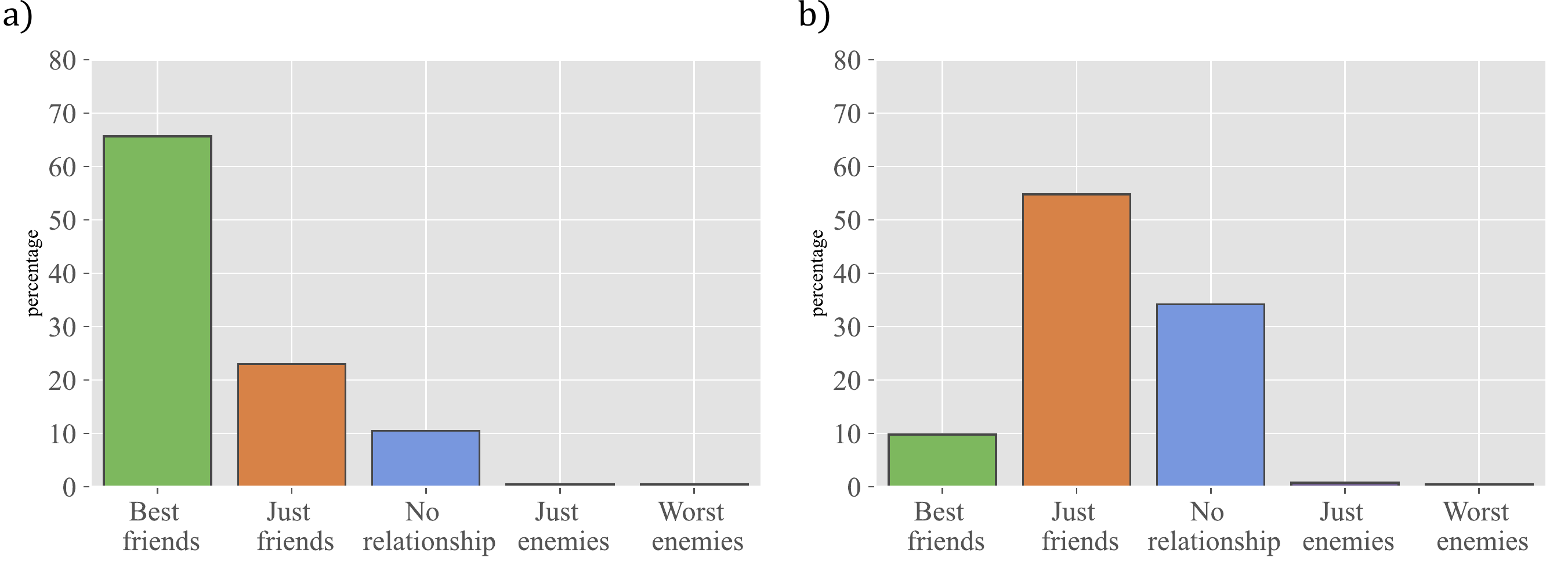}
 \caption{{\textbf{The evolution of friendships over time} - The figure represents the percentage of individuals that ended up in a given category in wave 5, when they were marked in wave 4 as: a) best friend, conditional probability $P(x,w_5|+2,w_4)$ or b) just friend, conditional probability $P(x,w_5|+1,w_4)$.}}
\label{fig_circles}
\end{figure}

\begin{figure}
\centering
\includegraphics[width=0.975\textwidth]{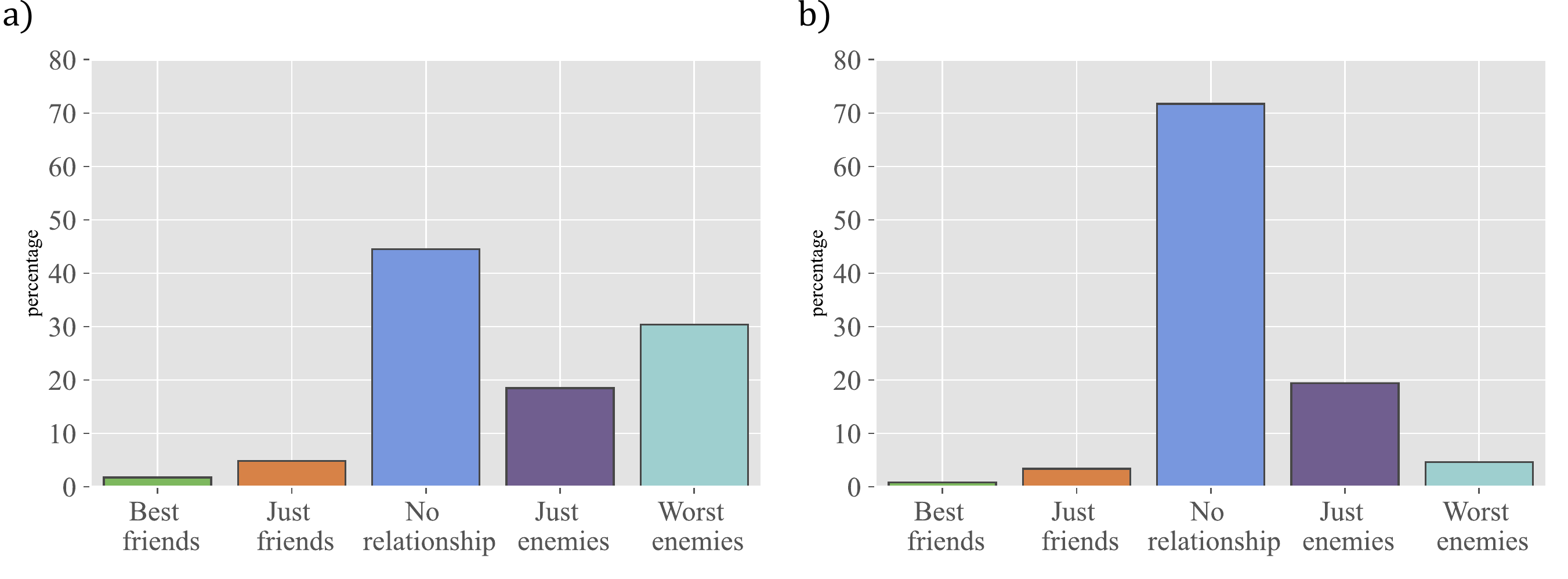}
 \caption{{\textbf{The evolution of enmities over time} - The figure represents the percentage of individuals that ended up in a given category in wave 5, when they were marked in wave 4 as: a) worst enemy, conditional probability $P(x,w_5|-2,w_4)$ or b) just enemy, conditional probability $P(x,w_5|-1,w_4)$.}}
\label{fig_circles2}
\end{figure}

On the other hand, we observe a larger degree of turnover than that reported in Ref.~\citen{roy2022}. This may be due to the fact that the latter study deals with data obtained from phone calls between adults; interestingly, even in that study participants aged 17--21 showed a larger turnover than those older than 21 years old (see also Ref.~\citen{tamarit2018}). Nonetheless, even Ref.~\citen{roy2022} reported differences between layers similar to what we find here. This points to the role of developmental issues in the evolution of the structure of personal friendship networks. Care has to be taken, though, because the phone data should capture the general structure of people's friendships, whether they are family, workmates, friends, etc. In contrast, here we are restricting the students to only their relationships within the school. In this respect, what we are seeing is that the Dunbar circle structure reproduces itself in each domain of relationships: a fraction of a person's cognitive capabilities are devoted to school relationships, and then from that limited capability the structure follows as Ref.~\citen{tamarit2018}. The more rapid turnover could be related to the smaller cognitive capacity devoted to the specific niche of school relationships. 

\bigskip

One possible confound that might influence the evolution of relationships is the distribution of students in the different classes. Generally speaking, some 70\% of the students' relationships are with other students in their same year, and of those, a majority are with students in their same class. In fact, among all potential relationships within the same class, approximately 50\% are actually reported, whereas less than 5\% of all potential relationships with students in different classes are actually realised (see Fig.~\ref{fig_percentage}{{a}}; see Sec.\~S6 in the Supplementary Information for a comparison of the total numbers of such friendships). The fact that this is an important factor can be seen also in Fig.~\ref{fig_percentage}{{b}}, where pairs of students are divided into groups according to whether they were in the same class in the two academic years included in our data (2020-2021 and 2021-2022) (hereafter referred to as S-S), in different classes both years (hereafter D-D), in the same class in the first year and in a different class in the second year (hereafter S-D), and vice versa (hereafter D-S). Then, Fig.~\ref{fig_percentage}{{b}} shows the percentage of relationships in each of these groups that were actually reported in each wave. Importantly, the change in the academic year between wave 2 and wave 3 students is associated with a reshuffling of the classes. This reflects in a decrease of relationships S-D, going from values close to 50\% to 25\% (orange bars in Fig.~\ref{fig_percentage}{{b}}), i.e., the separation led to the disappearance of half of the existing relationships. In contrast, the plot shows an increase in the percentage of relationships D-S (blue ones in Fig.~\ref{fig_percentage}{{b}}); in this case, the percentage rises from 15\% to almost 45\%, comparable to the starting point of the other group. This observation should be compared to the almost constant percentages of S-S and D-D relationships. This clearly shows that being in the same class is a very relevant driver for relationships to decay or start. In this regard, it is also interesting to note that when students are separated, the number of relationships that still remain in the second year is almost twice as large as relationships D-S in the first year, suggesting that it takes longer for relationships to disappear due to separation than to form upon becoming together.

\begin{figure}
\centering
\includegraphics[width=0.975\textwidth]{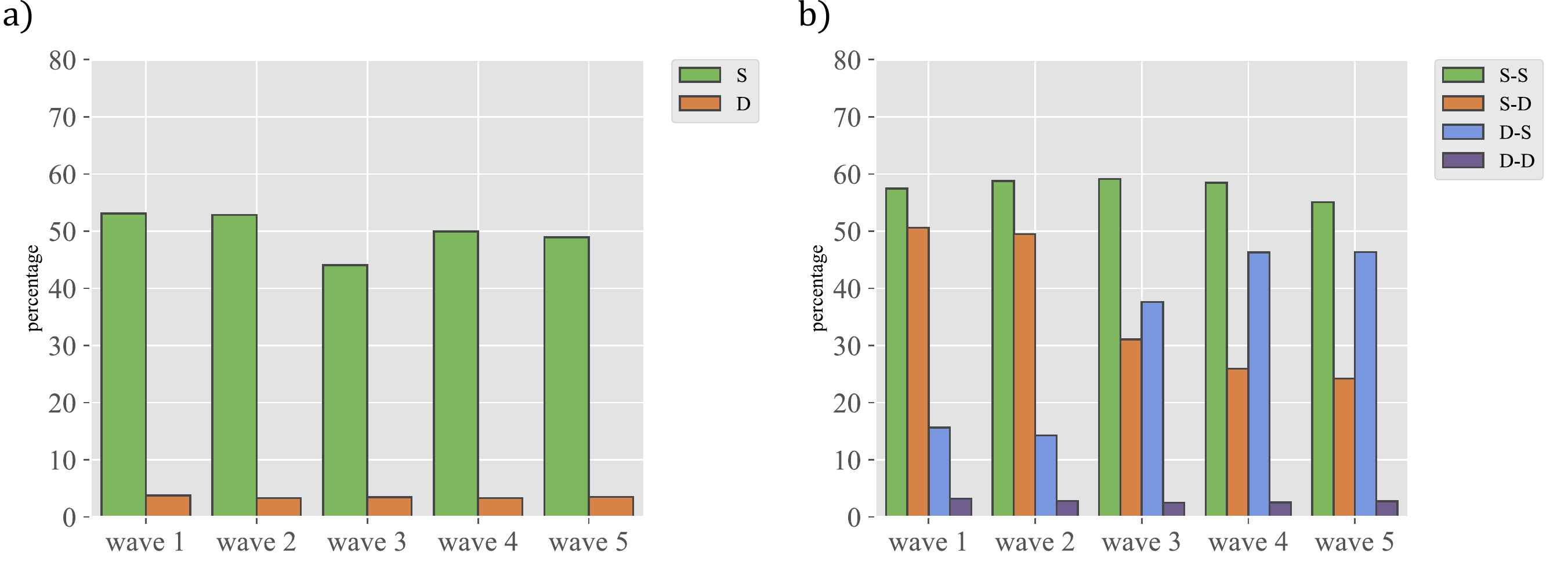}
\caption{{\textbf{The importance of the group for the existence of relationships} - The figure represents: a) percentage of relationships formed between individuals in the same group (S) versus those in different groups (D) relative to the total number of relations that might potentially form in each of the two cases and b) percentage of relationships between individuals that are in the same group both academic years (S-S), in the same group the first year but different the second (S-D), in a different group the first year and the same group the second (D-S) and different group both years (D-D), referred in each case to the total number of possible relationships. }}
\label{fig_percentage}
\end{figure}

Our longitudinal study also allows us to address the issue of the reciprocity of relationships. As shown in Fig.~\ref{fig_reciprocity}, the aggregate percentage of reciprocal relationships is remarkably similar in all five waves and close to 60\%. On the other hand, as shown in the right panel of Fig.~\ref{fig_reciprocity}, although this is also true for most individuals, there are quite a few cases in which reciprocity is very low. Figures in Sec.~S7 of the Supplementary Information show that the results do not significantly depend on the group, the gender, or the itinerary. { On the other hand, regarding enmities, the small number of them that has been reported translates into a very noisy distribution of the reciprocity values (see Sec.~S8 in the Supplementary Information), and no clear pattern can be distinguished.}

As reciprocity is also a property of relationships, it is worth considering their dependence on the personal characteristics of the individuals involved. Figure~\ref{fig_reciprocity2} shows the percentage of reciprocal links between individuals of the same gender, and also the percentages of the four types of temporal evolution discussed above. Regarding gender, the plot shows that homophilic links are generally more reciprocal, while mixed-gender links are less reciprocal. Interestingly, when mixed-gender links are not reciprocal, it is not due to a gender bias (cf.~Sec.~S7 in the Supplementary Information). We can also see that reciprocity is quite high in links between students that remain together the two academic years (S-S), is lowest for students that are always separate (D-D), while S-D and D-S links decrease or increase, respectively, in reciprocity in later waves. Similar results arise when looking at relationships in the same class or itinerary (cf.~Sec.~S7 in the Supplementary Information). Reciprocated friendships are also more stable, as are triangles formed only by positive relationships (in both cases, they are much more stable than any other combination). These results are shown in Fig.~\ref{fig_links}.

\begin{figure}
\centering
\includegraphics[width=0.475\linewidth]{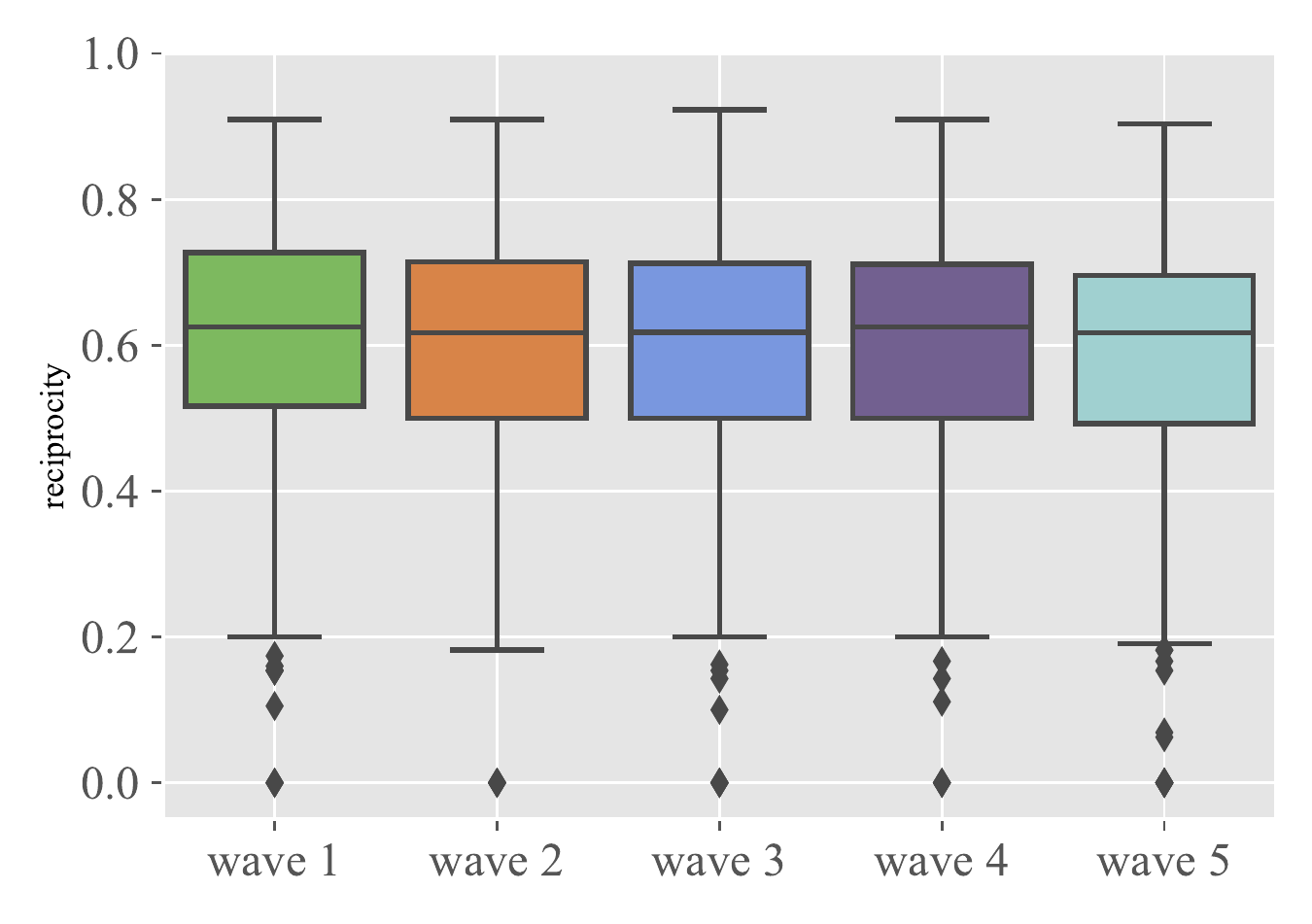} 
\caption{{\textbf{The evolution of reciprocity over time} - The figure represents boxplots that correspond to the distributions of individual reciprocal relationships in each wave. }}
\label{fig_reciprocity}
\end{figure}

\begin{figure}
\centering
\includegraphics[width=0.975\textwidth]{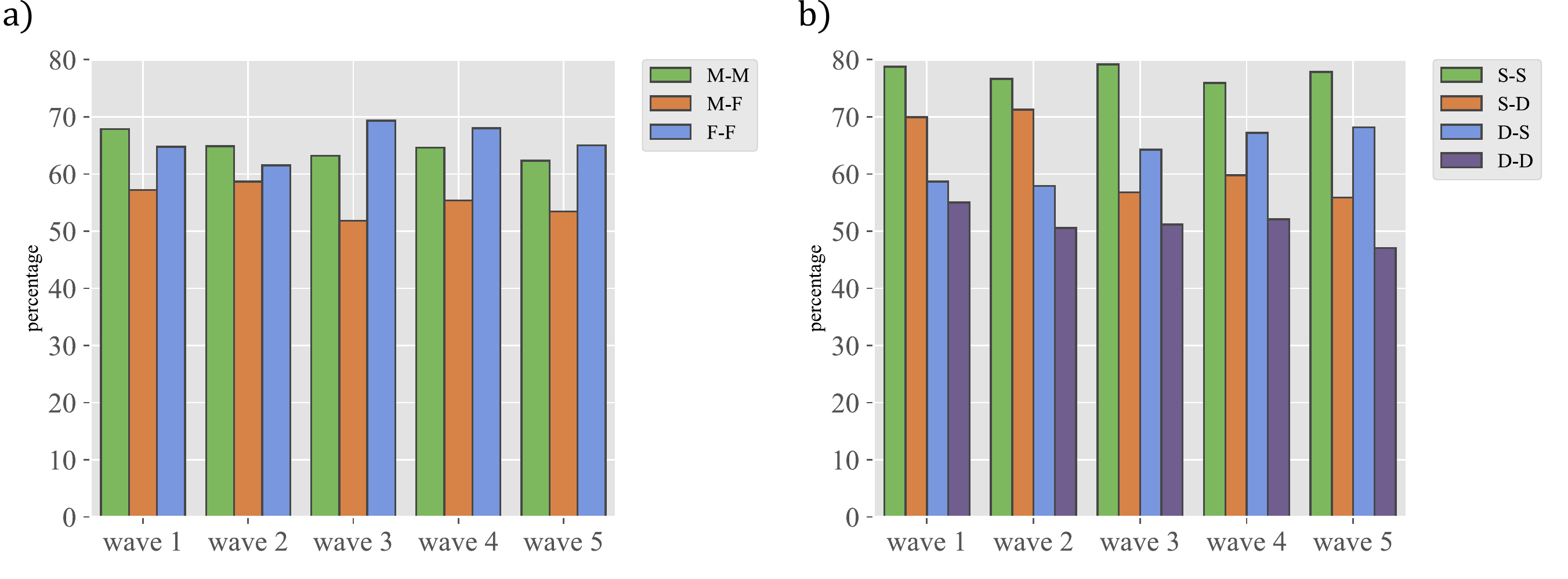}
\caption{{\textbf{The importance of the group for the reciprocity of relationships} - The figure represents: a) percentage of reciprocal links according to the individual's gender, M-M (male-male), M-F (male-female), F-F (female-female), and b) percentage of reciprocal links according to whether the pair of individuals are in the same or different class in consecutive years. }}
\label{fig_reciprocity2}
\end{figure}

\begin{figure}
\centering
\includegraphics[width=0.975\textwidth]{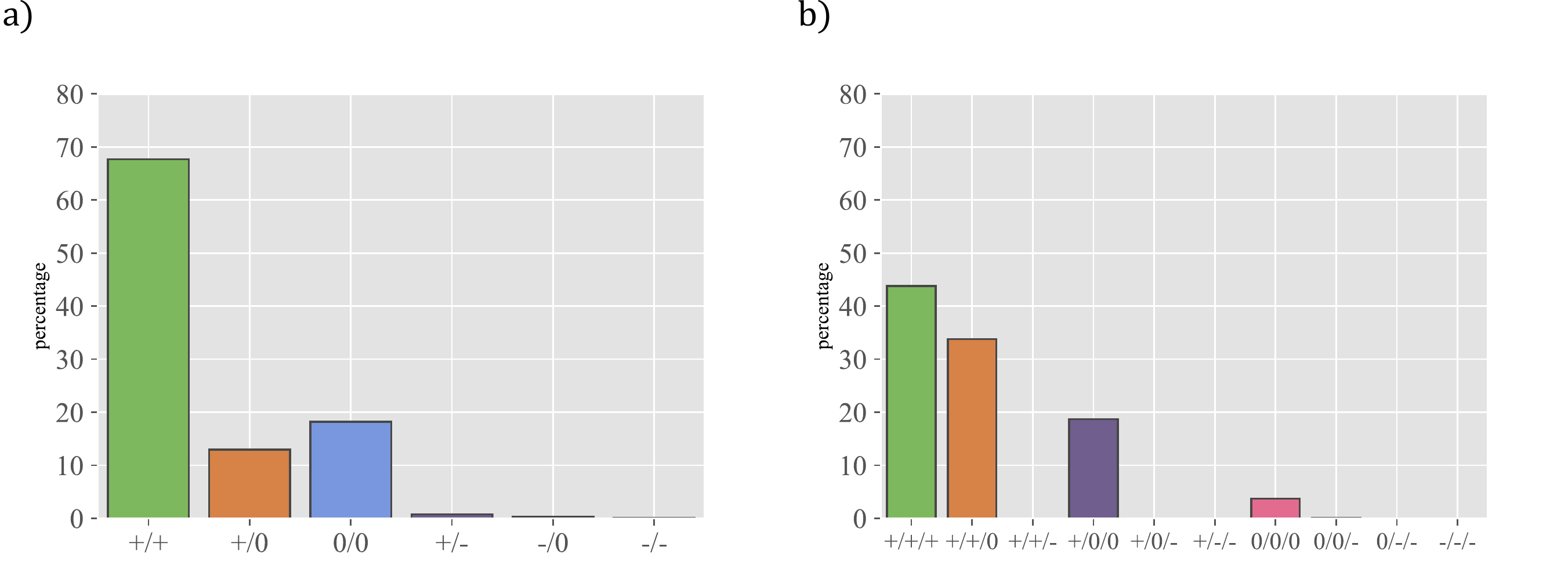}
\caption{{ \textbf{The evolution of reciprocal friendships over time} - The figure represents: a) percentage of pairs of links +/+ in wave 4 that end up in $x$/$y$ in wave 5, and b) percentage of reciprocal friendship triangles (type +/+/+) in wave 4 that end up in a triangle $x$/$y$/$z$ in wave 5, where the possible values are $x,y,z=+,-,0$.}}
\label{fig_links}
\end{figure}

\section*{Conclusions}

In this paper, we have studied the temporal evolution of relationships among 12-16 years old students that attend the same middle school. The study consisted of five waves of surveys during two consecutive academic years and included both positive and negative relationships. The number of students answering all five waves of the survey and considered in our analysis was 221.  In spite of what one could expect, we do not observe any signs of fatigue among the students, and their responses are remarkably similar in every wave, thus confirming the consistency of the data collection reported in Ref.~\citen{kucharski2018}. Indeed, the number of reported friends and best friends is quite constant in the five waves and it is so irrespective of  groups and ages, genders, itineraries, or being a ``repetidor''. Therefore, we have a very rich longitudinal dataset that allows a number of important issues to be addressed. 

The results from the surveys show clearly that friendships in the inner circle (best friends) are more stable than the rest of the friendships in the second circle (just friends). This observation provides further evidence of the key role of Dunbar's circles in the organisation of relationships, but also supports the idea that the intensity of the relationship in the first circle is higher than in the second one, apparently making them more stable. In contrast, enmities are few, much less frequent than friendships, and highly volatile, with many simply disappearing from one wave of the survey to the next. The different character of enmity and friendship networks has also been recently pointed out in Ref.\ \citen{sanchez23}. Note, however, that this does not mean that learning about enmities is irrelevant, as they have been shown to control the community structure of classes in Ref.~\citen{escribano2021}. This fact is therefore important for the daily dynamics of the class and as such, it is highly valuable information for the school management. 
    
Our study also points to the importance of being in the same class for forming and stabilising friendships. As we have discussed above, the strongest friendships arise among students who were in the same class during the two academic years we have studied. The change from being in the same class one year and separate the following one leads to the loss of a sizable fraction of friendships, which are then refocused on new classmates. Friendships among students that never shared class are much rarer in comparison. These observations suggest that we tend to have our relationship structure occupied at all times, as the friends who are lost because of the separation are replaced with new classmates. In addition, it also highlights the importance of frequent interaction in keeping or weakening relationships. 

A remark is in order concerning the comparison of these results with the preliminary ones reported in Ref.~\citen{escribano2021}. 
In terms of the total number of relations reported as friends and best friends, the numbers in Fig.~1 of Ref.~\citen{escribano2021} (notice that they refer to individual groups of the same year and not to whole years as reported here) are larger than those found here (cf.~Fig.~1 above). That study was done only with students from 1st year of ESO, and in the work reported here, they have answered the surveys again, now as students of 4th year of ESO, and their aggregate numbers of friendships are indeed different from those reported earlier. In contrast, students in this work who are in 1st year ESO report a number of friendships that is also smaller than that in Ref.~\citen{escribano2021} and similar to the rest of the classes in these surveys. Unfortunately, the students who took part in both studies cannot be connected because of an upgrade of the software, so we cannot make a proper longitudinal study encompassing the sample period that could explain the difference. In any event, the discrepancy is probably due to the presence of outliers in the study of Ref.~\citen{escribano2021} who have been removed in the present one. A few outliers report more than 100 relationships, thus biasing the mean to higher values. The much larger number of participants and waves in the present work makes the results much more reliable. In connection with this, we note that the distribution of slopes of fits to the temporal evolution of $\mu$ (see Sec.~S4 in the Supplementary Information) is relatively wide (much more so than the distribution of slopes for the evolution of the total number of friends), but we did not find any systematic correlation between the slopes and the year in which students were at the time of the waves. 
   
An interesting observation concerns the topic of reciprocal friendships, an important topic in view of its connection to performance\cite{vaquera2008,candia2022} and the success of behavioural interventions.\cite{almaatouq2016} Reciprocity is remarkably constant and for a majority of individuals, a percentage between 50\% and 70\% of their relationships are reciprocal. In general, gender homophilic relationships are slightly more reciprocal while male-female relationships tend to be less reciprocal, with both genders being equally responsible for this effect. Relationship reciprocity also shows the effect of group reshuffling and evolves in a manner similar to the friendships themselves. On the other hand, there are a few individuals whose reciprocity is very low, which could be an indicator of possible socialisation problems for those particular subjects, providing yet another valuable hint for the school management. In this regard, the results in Ref.\ \citen{burnett2015} show that reciprocity determines the resources allocated to different friendships, which is at the very root of their stability. In fact, these authors show that 17-year old students devote a lot of resources to reciprocated relations, while 14-year old students focus their effort on those they like irrespective of the reciprocation. In our study, we have 12- to 16-year old students, so the relevance of reciprocity as an indicator of possible relationship problems is higher for those as they may be misallocating their socializing efforts. 

{ Our focus has been on young teenagers, and a legitimate question to ask is whether their behaviour is representative of older teenagers and adults. Though a detailed answer is beyond the scope of the present study, we may note that the consistency in both the size and the relative temporal stability of the inner circle of friendships is in line with previous findings for both older teenagers \cite{roberts2011,saramaki2014} and adults.\cite{dunbar2020,roy2022} Indeed, the size of the inner core of best friends remains remarkably stable across the lifespan between the ages of 18 and 80.\cite{dunbar2016} This suggests that there is a tight cognitive constraint on the number of individuals who can be held in the innermost friendship circle at any one time, and that this remains relatively consistent across the lifespan despite the fact that the whole social network increases considerably in size between the early teenage years and adulthood.\cite{dunbar2016} The fact that the inner core circles of these young people's networks remain so stable in size, combined with the observation that network size in 18-year-olds is only around 50 people \cite{roberts2011} rather than the 150 normally associated with adults,\cite{dunbar2020} suggests that, over the course of childhood, social networks increase in size as a result of layers being successively added rather than by increasing the size of each layer while the number of layers themselves remains stable. This implies that, while there is a target size in adults that has evolutionary origins, this is achieved as a result of a learning process associated with developing social skills.\cite{henzi2007} Nonetheless, membership of the inner circle has a relatively higher turnover in the teenage years than it does in adults. Saram\"aki et al.,\cite{saramaki2014} for example, found that turnover in the network as a whole was around 30\% per annum in terms of movement in and out of the various circles. This may well reflect the fact that teenagers and young adults are sampling the options available locally in terms of friendship before finally settling down with more stable close friendships (see also Ref.~\citen{dunbar2016}).}

We end by summarising the big picture that can be inferred from this study. As already mentioned, everybody seems to have a predefined structure for their relationships, despite their frequent turnover. 
{
The results described so far---the stability of the number of relationships across waves and the higher turnover of the outer friendship layer ($+1$) compared to the inner circle ($+2$)---suggest picturing individuals as ``social atoms''. In this metaphor, layers play the role of atomic orbitals, whereas individuals act like electrons. Inner orbitals attract electrons more strongly than outer ones, so there is less turnover. Also, electrons may leave their orbitals for good, leaving a ``hole'' that is quickly filled by a new electron. Likewise, friends who leave the ego-network get replaced by new friends, so their average number remains constant.}
This suggests the possibility of studying the formation of social networks as a statistical-mechanical system in equilibrium, every relation having an associated ``binding energy''---the cost to remove the link. The question of whether one could then map this system into one of the available statistical-mechanical models of graph formation (e.g., those based on exponential random graphs \cite{strauss1986,park2004,escribano2022}) remains open. From a more philosophical point of view, one such mapping would imply that it is the ``total energy'' (the Hamiltonian, in the statistical-mechanical jargon) what describes a social system, rather than a graph. Graphs are volatile, and in constant change, whereas the energy uniquely determines the graph ensemble, of which any observed social network would be but a specific instance. One such mapping would open a big avenue to re-think social systems from a new perspective.


\section*{Methods}\label{methods}

{
\subsection*{School organizative structure}
}

In the academic year 2020-2021, due to COVID prevention measures, the structure of the school was different from the standard one. Students who were in 1st and 2nd year of ESO during the course 2020-2021 were divided into eight classes with some 15 students each. In the same year, students in the 3rd year of ESO were divided into five classes of some 25--30 students each, and within each class, they were split into two subgroups which, because of COVID, attended school physically on alternate days. In the academic year 2021-2022, these students advanced to the 2nd, 3rd and 4th year of ESO as already stated, and the school returned to a pre-COVID structure, i.e.~5, 5 and 4 classes with some 25 students in each year respectively, attending school physically on a daily basis. { Irrespective of the size, each class has a responsible teacher (``tutor'') and is an administrative unit.} In addition, there are two teaching pathways in this school, one that is taught mostly in English (except Spanish and Mathematics) and another that is taught mostly in Spanish (except Plastic Expression and Physical Education, which are taught in English). Approximately 40\% of the students take the English pathway (see Sec.~S1 in the Supplementary Information). 

\subsection*{Data collection}

Data were obtained from surveys conducted in the school involved in the study. The study was approved by the Ethics Committee of Universidad Carlos III de Madrid, and the surveys were subsequently carried out in accordance with the approved guidelines. Consent was obtained from the school which adopted this as a research project of its own and in turn got informed consent from the participants' parents. Students always participated voluntarily and signed informed consent prior to beginning the survey. { Participation was offered to all the students in the ESO courses, and for each wave there was a 90\% participation rate, with non participation due always to the fact that the student did not come to the school on that day.} The surveys were administered through a computer interface and included direct questions about their relationships as well as others related to personality traits. To elicit relationships, students could choose from a list of all the other ESO students in the school. For each individual in the school, we elicit friends, best friends, and light and strong enemies collected from each student as a list of student IDs and labelled as $\{+2,+1,-1,-2\}$. They will constitute the links of the personal networks. Note that each student provides this list, so we are extracting a directed network. 

{ Specifically, the computer administered questionnaire contained the following questions:
\begin{enumerate}
\item In what class (year and class; e.g., 1C, 2B) are you actually enrolled?
\item In what course were you enrolled last year?
\item You can now see the list of all the students in the school. Please mark those you have any relationship with by clicking very good relationship, good relationship, bad relationship or very bad relationship. Only one choice is possible. If you do not mark any option, it will be understood to mean that you do not have a relationship with the person. 
\end{enumerate}
This survey typically took 15 minutes to be filled by the students. We did not include further questions to prevent the study being too time consuming. Students were supervised by a school teacher during the whole process. We note that the survey also provides us with information about the variables "course" (i.e., 1 ESO, 2 ESO, etc), "repetidores" (if in the second question they answer with the same course) and itinerary (because the letter indicating the class corresponds to a specific one). Gender is obtained directly from the listings provided by the school and no question is needed. 
}

{
\subsection*{Outlier removal}
}

Outliers were removed according to the following criterion: students stating that they had more than 100 relationships and/or more than a 200\% change (upwards or downwards) in the number of reported relationships from one wave to the next were discarded from the sample. From the 285 students that answered in all five waves, we discarded 64 as outliers by this definition. 22 reported more than a hundred relationships, while 42 had too much variation between the number of answers in consecutive samples. This leaves us with a final sample of 
answers from 221 students for the five waves. { We emphasize that even if this looks like a drastic reduction of the sample, the results obtained when keeping these outliers are essentially the same as those reported here. As a check, we reran all the analyses with the full sample (i.e. including the outliers): none of the results differed from those reported above (see Sec.~S9 in the Supplementary Information). Nonetheless, we prefer to retain the main analyses since these are more conservative and reliable.}


\section*{Acknowledgements}

This work has been supported in part by grant PID2022-141802NB-I00 (BASIC) funded by MCIN/AEI  and by “ERDF A way of making Europe”.
\section*{Author contributions statement}

D.E., J.A.C. and A.S.\ conceived the research, D.E., F.J.L. and A.S. collected the data, D.E. analysed the data, and all authors discussed and interpreted the results and wrote the manuscript.

\section*{Data availability}

The datasets generated and/or analysed during the current study are not publicly available due to continuing use in parallel research projects from this group as well as from other collaborating groups, but are available from the corresponding author on reasonable request.

\section*{Ethics declarations}

\subsection*{Competing interests}

The authors declare no competing interests.


\begin{thebibliography}{}

\bibitem{dunbar2020} Dunbar, R.I.M. (2020). Structure and function in human and primate social networks: Implications for diffusion, network stability and health. Proc. R. Soc. Lond. 476A: 20200446.

\bibitem{saramaki2014} Saramäki, J., Leicht, E., López, E., Roberts, S.B.G., Reed-Tsochas, F. \& Dunbar, R.I.M. (2014). The persistence of social signatures in human communication. PNAS 111: 942-947. 

\bibitem{roberts2009}
Roberts, S.B.G., Dunbar, R.I.M., Pollet, T. \& Kuppens, T. (2009). Exploring variations in active network size: constraints and ego characteristics. Social Networks 31: 138-146.

\bibitem{roberts2011}
Roberts, S.B.G. \& Dunbar, R.I.M. (2011). Communication in social networks: effects of kinship, network size and emotional closeness. Pers. Relationships 18: 439-452.

\bibitem{shrum2008}
Shrum, W., Cheek, N. H., \& Hunter, S. M. (1988). Friendship in school: Gender and racial homophily. Sociology of Education 61, 227–239.

\bibitem{miritello2013}
Miritello, G., Moro, E., Lara, R., Martínez-López, R., Belchamber, J., Roberts, S.B.G. \& Dunbar, R.:I.M. (2013). Time as a limited resource: communication strategy in mobile phone networks. Social Networks 35: 89-95

\bibitem{sutcliffe2012}
Sutcliffe, A.J., Dunbar, R.I..M., Binder, J. \& Arrow, H. (2012). Relationships and the social brain: integrating psychological and evolutionary perspectives [with commentaries]. Brit. J. Psychol. 103: 149-

\bibitem{tamarit2018}
Tamarit, I., Cuesta, J., Dunbar, R. \& Sánchez, A. (2018). Cognitive resource allocation determines the organisation of personal networks. PNAS 115: 1719233115.

\bibitem{escribano2021} Escribano, D., Doldán-Martelli, V., Lapuente, F.J. et al. Evolution of social relationships between first-year students at middle school: from cliques to circles. Sci Rep 11, 11694 (2021).

{\bibitem{escribano2022b} Escribano, D., Doldán-Martelli, V., Cronin, K.A. et al. Chimpanzees organize their social relationships like humans. Sci Rep 12, 16641 (2022).}

{\bibitem{wrzus2013} Wrzus, C., H\"anel, M., Wagner, J. \& Neyer, F.J.
Social network changes and life events across the life span: a meta-analysis. Psychol Bull
139, 53-80 (2013).}

{\bibitem{kunal2016} Bhattacharya, K., Ghosh, A., Monsivais, D. et al. Sex differences in social focus across the life cycle in humans. R Soc Open Sci. 3, 160097 (2016).
}

{\bibitem{sekara2016}  Sekara, V.,  Stopczynski, A., \& Lehmann, S. Fundamental structures of dynamic social networks. Proc Natl Acad Sci USA 113, 9977-9982
 (2016).}








\bibitem{kucharski2018} Kucharski AJ, {\em et al}. (2018) Structure and consistency of self-reported social contact networks in British secondary schools. PLOS ONE {\bf 13}, e0200090.

\bibitem{smirnov2017} Smirnov, I. \& Thurner, S. (2017) Formation of homophily in academic performance: Students change their friends rather than performance. PLOS ONE {\bf 12}, e0183473. 

\bibitem{south2004} South, S.J. \& Haynie, D.L. (2004) Friendship Networks of Mobile Adolescents. Social Forces {\bf 83},  315--350.

\bibitem{sanchez23} Sánchez-Espinosa, D. B., Hernández-Ramírez, E., \& del Castillo-Mussot, M. (2023). Popularity and Entropy in Friendship and Enmity Networks in Classrooms. Entropy 25, 971.

\bibitem{kenna2017} Yose, J., Kenna, R., MacCarron, M., \& MacCarron, P. (2018). Network analysis of the Viking age in Ireland as portrayed in {\em Cogadh Gaedhel re Gallaibh}. Royal Society Open Science, 5(1), 171024.

\bibitem{roy2022} Roy, C., Battacharya, K. Dunbar, R.I.M., and Kaski, K. Turnover in close friendships. Sci Rep 12, 11018 (2022).

{
\bibitem{vaquera2008} Vaquera, E. \& and  Kao, G. Do You Like Me as Much as I Like You? Friendship Reciprocity and Its Effects on School Outcomes among Adolescents. Soc Sci Res. 37, 55–72 (2008).
}
\bibitem{candia2022} Candia, C., {\em et al.} (2022) Reciprocity heightens academic performance in elementary school students.
Heliyon {\bf 8} e11916. 

\bibitem{almaatouq2016} Almaatouq, A., Radaelli, L., Pentland, A., \& Shmueli, E. (2016) Are You Your Friends’ Friend? Poor Perception of Friendship Ties Limits the Ability to Promote Behavioral Change. PLoS ONE {\bf 11}, e0151588.

{\bibitem{burnett2015} Burnett Heyes, S., Jih, Y.-R., Block, P. et al. 
Relationship Reciprocation Modulates Resource Allocation in Adolescent Social Networks: Developmental Effects. Child Development 86,  1489–1506 (2015).}



\bibitem{strauss1986} Strauss, D. (1986) On a General Class of Models for Interaction. SIAM Rev. {\bf 4}, 513--527.

\bibitem{park2004} Park, J. and Newman, M. E. J. (2004) Statistical mechanics of networks.
Phys. Rev. E {\bf 70}, 066117.

\bibitem{escribano2022} Escribano, D., and Cuesta, J. A. (2022) Free-energy density functional for Strauss’s model of transitive networks. Phys. Rev. E {\bf 106}, 054305.

\bibitem{dunbar2016} Dunbar, R.I.M. (2016). Do online social media cut through the constraints that limit the size
of offline social networks? Royal Society Open Science 3: 150292.

\bibitem{henzi2007} Henzi, S.P., de Sousa Pereira, L., Hawker-Bond, D., Stiller, J., Dunbar, R.I.M. and Barrett, L.
(2007). Look who’s talking: developmental trends in the size of conversational cliques.
Evolution and Human Behavior 28: 66-74.

\end{thebibliography}
\end{document}